\documentstyle[twocolumn, eqsecnum,aps,graphicx]{revtex}
\begin{document}

\title{\bf Non-adiabatic Landau-Zener transitions in low
spin molecular magnet V$_{15}$}

\author{I. Chiorescu$^{a\sharp}$, W. Wernsdorfer$^a$, A. M\"uller$^b$, H.
B\"ogge$^b$ and B. Barbara $^a$.}

\address{$^a$ Laboratoire de Magn\'etisme Louis N\'eel, CNRS, BP 166,
38042-Grenoble, France.\\
 $^b$ Fak\"ultat f\"ur Chemie, Universit\"at Bielefeld, D-33501 Bielefeld,
Germany. }

\date{\today}

\maketitle

\begin{abstract}
The V$_{15}$ polyoxovanadate molecule is made of 15 spins $1/2$ with
antiferromagnetic couplings. It belongs to the class of molecules with very
large Hilbert space dimension (2$^{15}$ in V$_{15}$, 10$^8$ in Mn$_{12}$-ac).
It is a low spin/large molecule with spin $S=1/2$. Contrary
large spins/large molecules of the Mn$_{12}$-ac type, V$_{15}$ has no energy
barrier against spin rotation. Magnetization measurements have
been performed and despite the absence of a barrier, magnetic hysteresis is
observed over a timescale of several seconds. This new phenomenon
characterized by a ``butterfly'' hysteresis loop is due to the effect of the
environment on the quantum rotation of the entangled 15 spins of the molecule,
in which the phonon density of states is not at its equilibrium (phonon
bottleneck). 

\noindent keywords: magnetic quantum effects, relaxation phenomena,
dissipative two-level system.

\end{abstract}

\textbf{1. Introduction}\\

The recent evidence for quantum tunneling of the magnetization in
big molecules with large spin $S=10$ \cite{NatFriedJMMM} allows a link to be
made between mesoscopic physics and magnetism \cite{BB}. In these systems,
energy barriers are high and tunnel splittings between symmetric and
antisymmetric quantum states are exponentially small (10$^{-7}-10^{-11}$~K).
As a consequence, quantum relaxation is slow and only related to the spin
baths \cite{GargStamp,Stamp0,ITBBEMCJT}. Resonant phonon transitions are irrelevant,
unless between states at different energies \cite{BoutronVillainLoss} or in
the presence of a transverse field large enough to create a tunnel splitting
of the order of the temperature energy scale \cite{Belessa}.  Here we show
that the reversal of an ensemble of non-interacting spins $S=1/2$ is
irreversible as a consequence of different couplings to the environment. Such
an effect gives rise in our system to a  ``butterfly'' hysteresis loop. Spin
rotation in V$_{15}$ may be viewed as an experimental realization of the
theoretical problem of a 2-level system with dissipation
\cite{LeggettGrifoni,PRLV15}. Spin-phonon transitions take place between the
two splitted ground states at well defined energies $\hbar\omega=\Delta_H$,
where $\Delta_H$ is the field dependent energy separation of the two levels.
The width $\Delta\omega$ of such transitions being relatively narrow, the
number of available phonon states at energy $\hbar\omega$ is much smaller than
the number of molecules and a hole must be burned in the spin-phonon density of
states \cite{A&B,VanVleckStev}. This hole, moving with the sweeping field,
leads to a non-equilibrium spin-phonon density of states, giving in the
V$_{15}$ complex the observed phonon bottleneck and hysteresis . A phonon
``bottleneck plateau'', identified on the measured hysteresis loop, gives a
characteristic butterfly shape. These results are corroborated with numerical
calculations based on this model.\\

\textbf{2. The V$_{15}$ magnetic moment evolution in low sweeping rates.}\\

\begin{figure}
\includegraphics[width=8.1cm]{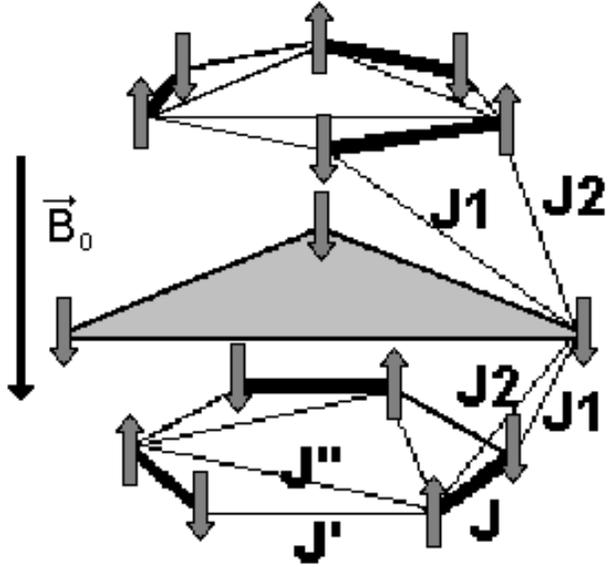}
\caption{The geometry and the spin coupling scheme in V$_{15}$ complex. There
are five antiferromagnetic exchange constants leading to a ground state $S=1/2$
in fields below 2.8~T and $S=3/2$ for greater fields (here such a field is
pointing down). }
 \label{molecule}
\end{figure}

The V$_{15}$ complex of formula  $K_6[V_{15}^{IV}As_6O_{42}(H_2O)]\cdot8H_2O$
was prepared as described in  \cite{Muller,GattMol,GattNature}. It is made of
a lattice of molecules with  fifteen V$^{IV}$ ions of spin $S=1/2$, placed in
a quasi-spherical layered  structure formed of a triangle, sandwiched by two
hexagons  (Fig.~\ref{molecule}). All the interactions being antiferromagnetic,
the  resultant (collective spin) is $S=1/2$. The symmetry is trigonal (space 
group $R\bar3c$, $a=14.029$~\AA, $\alpha=79.26^{\circ}$, $V=2632$~\AA$^ 3$). 
The unit-cell contains two V$_{15}$ clusters and it is large enough so that 
dipolar interactions between spin 1/2 molecules are negligible (few mK). 
Each hexagon contains three pairs of strongly coupled spins ($J\cong -800$~K)
and each spin at a corner of the inner triangle is coupled to two of  those
pairs (one belonging to the upper hexagon and one belonging to the  lower
hexagon ($J'\cong J_1\cong -150$~K, $J''\cong J_2\cong -300$~K 
\cite{GattMol}). Thus, the V$_{15}$ molecule can be seen as formed by three 
groups of five V$^{IV}$ ions with resultant spin 1/2 and assembled on each 
corner of the inner triangle. These three spins 1/2 interact to each others 
through two main paths, one passing by the upper hexagon and one passing by 
the lower one. This is a typical example of frustrated molecule,  where the
exchange $J_0$ between the spins 1/2, is much  smaller than the
exchange interactions between two spins. 

Two types of  single crystal magnetization measurements have been
performed: (i) characterization of the general thermodynamical properties of
the system at equilibrium and (ii) study of the non-equilibrium to equilibrium
transition.

\begin{figure}
\begin{center}\includegraphics[width=8.1cm]{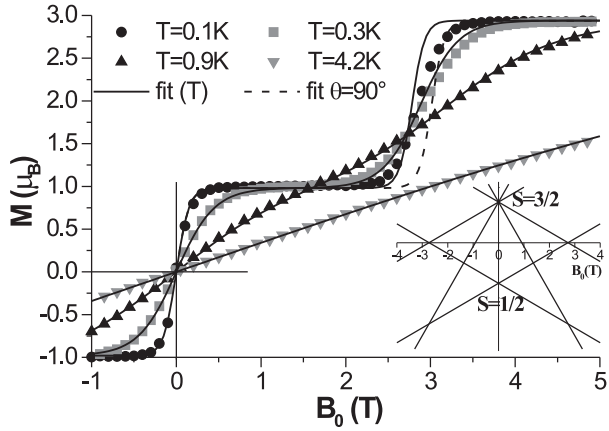}\end{center}

\caption{The agreement between the calculated curve at thermodynamical
equilibrium (line) and the experimental magnetization data is quite good for
an isotropic exchange $J_0=-2.445$~K. For $J_X=J_Y=-2.75$~K, $J_Z=-2.2$~K and
the field applied in the triangle plane (dashed line) one observes that the
calculated second flip is slightly displaced but its slope is still greater
than the experimental one. In inset are represented the energy levels vs.
applied field in the effective triangle approach. When the applied field
reaches $-3J_0/(2g\mu_B)$ the ground state switch from $S=1/2$ to $S=3/2$.} 
\label{jump}
\end{figure}

In the first case a small dilution refrigerator allowing measurements above
0.1~K, was inserted in an extraction magnetometer providing fields up to 16~T,
with low sweeping rates. Below 0.9~K we observed one jump in zero field and
two others at $B_1\cong\pm$ 2.8~T (Fig.~\ref{jump}). They correspond to the
$|1/2,-1/2\rangle\leftrightarrow|1/2,1/2\rangle$ and $|1/2,1/2\rangle
\leftrightarrow |3/2,3/2\rangle$ spin transitions, with respective saturations
at $0.50\pm 0.02~\mu_B$ and $2.95\pm 0.02~\mu_B$ per V$_{15}$ molecule, in
agreement with the level scheme given in Fig.~\ref{jump} inset. For all
temperatures the magnetization curves are reversible and do not show any
anisotropy, showing that eventual energy barrier preventing spin reversal must
be quite small (less than 50~mK).

The magnetization curves measured at equilibrium (Fig.~\ref{jump}) are fitted
using the Heisenberg Hamiltonian for three frustrated spins $S=1/2$:
\begin{eqnarray}
H=-\!\!\!\!\sum_{\alpha=X,Y,Z}J_{\alpha}(S_{\alpha 1}S_{\alpha 2}+S_{\alpha
2}S_{\alpha 3}+S_{\alpha 3}S_{\alpha1})  \nonumber \\ 
-g\mu_B\vec{B_0}(\vec{S}_1+\vec{S}_2+\vec{S}_3)
\label{Heis} 
\end{eqnarray}
with $S_1=S_2=S_3=1/2$, $J_X=J_Y<0, J_Z<0$ (antiferromagnetic coupling),
$g=2$, $B_0$ the applied field and $\mu_B$ the Bohr magneton. In the isotropic
case ($J_{X,Y,Z}=J_0$) the configuration $S=3/2$ becomes favourable against the
$S=1/2$ one for $B_1=-3J_0/(2g\mu_B)$. The measured value $B_1\cong 2.8$~T
gives $J_0\cong -2.5$~K. The comparizon of the calculated curves
($S_{Zi}=\langle i|S_{Z1}+S_{Z2}+S_{Z3}|i\rangle$ averaged over the different
eigenstates $E_i, |i\rangle, i=1..8$) with the measured ones is quite good at
any temperature. The width of the transition at $B_1\cong \pm$2.8~T is
nevertheless broader in the experiments (Fig.~\ref{jump}). This broadening of
about 0.7~T cannot be inferred to dipolar or hyperfine field distributions
(about 1~mT and 40~mT respectively). The observed anisotropy of the electronic
g-factor ($g_a$=$g_b$=1.95 and $g_c$=1.98) is also too small\cite{GattNature}.
Antisymmetrical Dzyaloshinsky-Moriya interactions (\cite{Dobro,ICM} and ref.
therein) allowed by symmetry and coupling the states $S=1/2$ and $S=3/2$,
could explain this broadening.

In low fields, susceptibility measurements give the following effective
paramagnetic moments: $\mu_{eff}=g\sqrt{S(S+1)}=1.75\pm0.02~\mu_B$
corresponding to $S=1/2$ below 0.5~K,  and $\mu_{eff}=3\pm0.02~\mu_B$
corresponding to 3 independent spins $S=1/2$ above 100~K. It confirms clearly
that the ground state of the molecule is $S=1/2$ below 0.5~K.\\

\textbf{3. Spin-phonon coupling in V$_{15}$ in low fields and high sweeping
rates: a dissipative spin 1/2 two-level system.}\\

 We showed that, between -2.8~T and 2.8~T, the total spin
of this molecule is $S=1/2$. Such a small spin leads to vanishingly small
energy barrier and relatively large splitting in zero-field ($\sim10^{-2}$~K).
Resonant spin-phonon transitions between the symmetric and antisymmetric
states at resonance are then possible in sub-Kelvin experiments, contrary to
the case of big molecules in zero field, where these states are much too
close. 

\begin{figure}
\begin{center}\includegraphics[width=8cm]{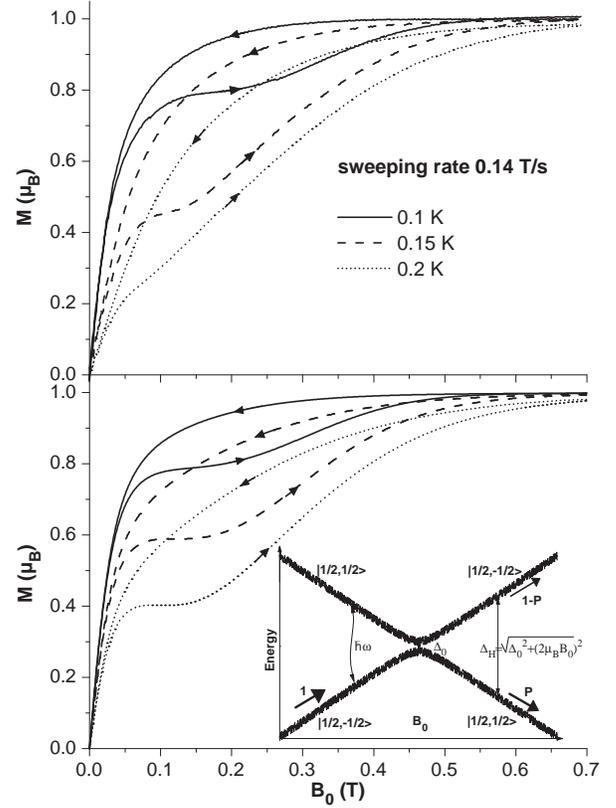}\end{center}

\caption{Measured (\textbf{a}$-$\emph{top}) and calculated
(\textbf{b}$-$\emph{bottom}) hysteresis loops for three temperatures and for
a given field sweeping rate 0.14~T/s. The plateau is more pronounced at low
T.  The inset is  a schematic representation of a two-level system $S_Z=\pm1/2$
with repulsion due to non-diagonal matrix elements. In a swept field the
switching probability $P$ is given by the Landau-Zener formulae (see text).
The two levels are broadened by the hyperfine fields and the absorption
or the emission of phonons can switch the polarization state of spins.}
\label{3Temps} \end{figure}

\begin{figure}
\begin{center}\includegraphics[width=8cm]{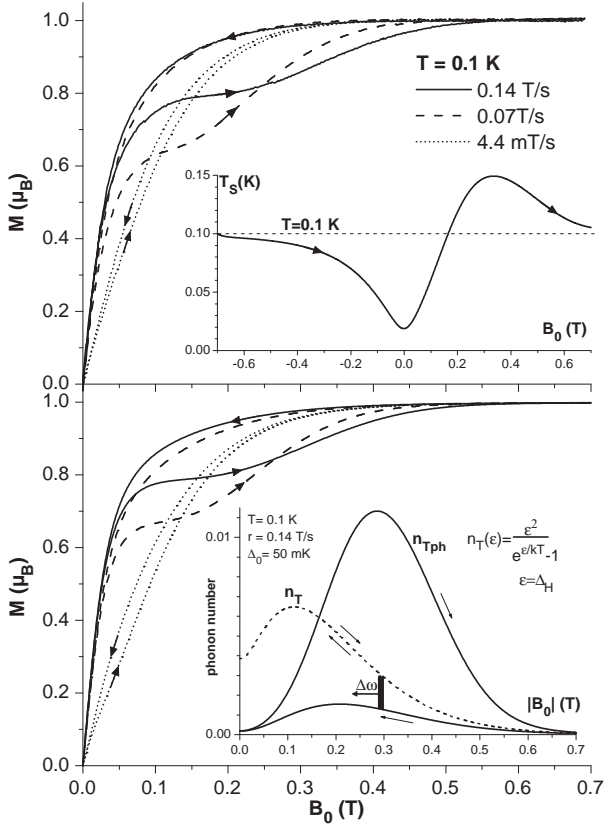}\end{center}

\caption{Measured (\textbf{a}$-$\emph{top}) and calculated
(\textbf{b}$-$\emph{bottom}) hysteresis loops for three field sweeping rates
at $T=0.1$~K. The observed plateau is more pronounced at high sweeping rate.
The equilibrium curve can be approximated by the median
of the two branches of the low sweeping rate hysteresis loop  (dotted curve). 
In the \emph{top} inset is plotted the spin and phonon temperature $T_S=
T_{ph}$ for $T=0.1$~K and $r=0.14$~T/s, when the field is swept from negative
values. $T_S$ decreases until zero-field and then increases linearly within the
plateau region. Then it overpasses the bath temperature to finally reach the
equilibrium. In the \emph{bottom} inset the calculated number of phonons with
$\hbar\omega=\Delta_H$ is plotted vs. the sweeping field modulus (note the
arrows) at equilibrium ($T_{ph}=T_S=T$, dashed line) and
out-of-equilibrium ($n_{T_{ph}}=n_{T=T_S}$, $r=0.14$~T/s, black line). The
difference between the two curves (thick segment $\Delta\omega$) suggests the
moving hole in the phonon distribution, while their intersection gives the
plateau intercept of the equilibrium magnetization curve.}

\label{3rates} \end{figure}

The sensitivity and time resolution of the micro-SQUID magnetometer
\cite{PRLFeBa} allowed to study very small V$_{15}$ crystals in good contact
with the thermal bath.  Down to 50~mK non-equilibrium behavior was
nevertheless observed at fast sweeping rate (up to $dB_0/dt=0.7$~T/s). 
Few hysteresis loops are represented Fig.~\ref{3Temps}a and Fig.~\ref{3rates}a
(only the positive part, the other one being rigorously symmetrical). When
the field increases, coming from negative values, the magnetization passes
through the origin of the coordinates, reaches a plateau and then approaches
the saturation. This leads to a winged hysteresis loop characterized by the
absence of irreversibility near zero field. The wings depend sensitively on
temperature and sweeping field rate. As an example, in Fig.~\ref{3Temps} where
three hysteresis loops are presented (T=0.1~K, 0.15~K, 0.2~K) for a given
sweeping rate (0.14~T/s), the plateau is higher and more pronounced at low
temperature. The same tendency is observed at given temperature and faster
sweeping rate (Fig.~\ref{3rates}). At a given temperature, the equilibrium
magnetization curve can be approximated by the median of the two branches of a
low sweeping rate hysteresis loop (e.g. in Fig.~\ref{3rates} for
$dB_0/dt=4.4$~mT/s). When compared to the equilibrium curve,  a given
magnetization curve shows that (i) in negative fields the system is colder
than the bath, (ii) the spin system remains colder until the magnetization
curve intersepts the equilibrium curve, (iii) after the intersept the spin
temperature overpass the bath temperature, (iv) at sufficiently high fields
(about 0.5~T) the system  reaches the equilibrium.    

In order to interpret this magnetic behavior of the  V$_{15}$ molecules, we
will analyse how the level occupation numbers vary in this two level system
(see Fig.~\ref{3rates}a inset) when sweeping an external field. In the
bare Landau-Zener (LZ) model \cite{Zener,MiyaKaya} the probability
for the $|1/2,-1/2\rangle\leftrightarrow|1/2,1/2\rangle$ transition is:
\begin{equation}
P=1-\exp\left(-\frac{\pi\Delta_0^2}{4\hbar\mu_Br}\right).
\label{PLZ} 
\end{equation}
 Taking the typical value
$r=0.1$~T/s and the zero-field splitting $\Delta_0\cong 0.05$~K
\cite{hyp,Carter}, one gets a ground state switching probability very
close to unity: the spin occupancy of the two level system is not modified
when the field is reversed. In such an LZ  transition, the plateaus of
Fig.~\ref{3rates} should decrease if the sweeping rate increases, which is
contrary to the experiments. Not surprizingly the $S=1/2$ spin system is not
isolated:  LZ transitions are non-adiabatic due to spin couplings to the
environment.

 	The mark of the V$_{15}$ system is that this coupling is acting also
near zero-field because $\hbar\omega\approx\Delta_0$ is of the order of the
bath temperature. The spin temperature $T_S$ is such that  $n_1'/n_2=
\exp(\Delta_H/k_BT_S)$, where $\Delta_H$ is the two levels field-dependent
separation, and $n_{1,2}$($n_{1,2eq}$) the out of equilibrium (equilibrium)
level occupation numbers. In the  magnetization curves at 0.1~K
(Fig.~\ref{3Temps}a, \ref{3rates}a), the spin temperature is significantly
lower than the bath temperature $T$ ($n_{1}>n_{1eq}$, $T_S<T$) between $-0.3$~T
(when the magnetization curve departs from the equilibrium one) and 0.15~T
(the field at which the magnetization curve intersects the equilibrium one).
	After this intersept $T_S$ is larger than the bath temperature
($n_{1}<n_{1eq}$, $T_S>T$), and at sufficiently high fields (about 0.5~T) it
reaches the equilibrium value ($n_{1}=n_{1eq}$, $T_S=T$). Note that the
magnetization curves measured between $-0.7$~T and 0.02~T at fast sweeping
rates (0.07 and 0.14~T/s) are nearly the same, suggesting weak exchange with
the bath, i.e. nearly adiabatic demagnetization. 

Indeed, as we will show below, the silver sample holder is not sufficient to
maintain the phonon temperature equal to the temperature of the bath. This is
because in V$_{15}$ below 0.5 K, the heat capacity of the phonons $C_{ph}$ is
very much smaller than that of the spins $C_S$, so that the energy exchanged
between spins and phonons will very rapidly adjust the phonons temperature
$T_{ph}$ to the spin one $T_S$. Thus, we can see the spin system and the
phonons as a single coupled system (quantum non-adiabatic LZ transitions) in
weak exchange with the external bath (thermodynamical adiabatic
demagnetization).  The spins energy is transfered from the spins only to
those phonons with $\hbar\omega=\Delta_H$ (within the resonance line width).
The number of such lattice modes being much smaller than the number of spins,
energy transfer between the phonons and the sample holder must be very
difficult, a phenomenon known as the phonon bottleneck \cite{VanVleckStev}.
Following \cite{A&B}, the number of phonons per molecule available for such
resonant transitions is $n_{T}=\int_{\Delta\omega}\sigma (\omega)
d\omega/(\exp(\hbar\omega/kT)-1)$, where $\sigma
(\omega)d\omega=3V\omega^2d\omega/(2\pi^2v^3)=$ number of phonon modes between
$\omega$ and $\omega+d\omega$ per  molecule of volume $V$, $v$ is the phonon
velocity and $\Delta\omega$ is the transition linewidth due to fast hyperfine
field fluctuations (they broden both energy levels). Taking the typical values
$v\approx 3000$~m/s, $T\approx10^{-1}$~K and $\Delta\omega\approx 10^2$~MHz
\cite{Stamp} we find $n_{T}$  of the order of $\approx10^{-6}$ to $10^{-8}$
phonons/molecule. Such a small number of phonons is very rapidly absorbed,
burning a hole of width $\Delta\omega$ in the spin-phonon density of states at
the energy $\hbar\omega=\Delta_H$ \cite{VanVleckStev}. If this spin-phonon
density of states does not equilibrate fast enough, the hole must persist and
move with the sweeping field, leading to a phonon bottleneck.\\

\textbf{4. Quantitative approach and numerical calculations.}\\

Now this description will be made quantitative. For a given splitting
$\Delta_H$, the time evolution of the two levels populations $n_{1,2}$ and of
the phonon numbers $n_{T_{ph}}$ at $T_{ph}$ obeys the set of two differential
equations \cite{A&B}: 
\begin{equation}
\left\{\begin{array}{l}
-\dot n_1=\dot n_2=P_{12}n_1-P_{21}n_2 \\
\dot n_{T_{ph}}=-(n_{T_{ph}}-n_T)/\tau_{ph}-P_{12}n_1+P_{21}n_2,
\end{array}
\right.
\label{syst1}
\end{equation}
where $P_{12,21}$ are the transition probabilities between the two levels
(they are themselves linear functions of $n_{T_{ph}}$) and
$\tau_{ph}\approx L/2v$ is the phonon-bath relaxation time ($L$ is the sample
size). Using the notations $x=(n_1-n_2)/(n_{1eq}-n_{2eq}),
y=(n_{T_{ph}}-n_T)/(n_T+n/2)$ with $n=\int_{\Delta\omega}\sigma (\omega)
d\omega$ we get: 
\begin{equation}
\left\{\begin{array}{l}
\dot x=(1-x-xy)/\tau_1 \\
\dot y=-y/\tau_{ph}+b\dot x,
\end{array}
\right.
\label{syst2}
\end{equation}
where $b=C_S/C_{ph}$ and $1/\tau_1=P_{12}+P_{21}$
the direct spin-phonon relaxation time. By solving numerically this system
for typical values, e.g. $\tau_1=10^{-2}$~s, $\tau_{ph}<10^{-6}$~s,  $b>10^5$,
we can see that $T_{ph}\rightarrow T_S\not=T$ (phonon bottleneck) very
rapidly, as expected.This leads to  $y=1/x-1$ and the second equation of the
differential system becomes $\dot x=(x-x^2)/(1+bx^2)/\tau_{ph}$. In the limit
$b>>1$ (in our case $b\approx10^8-10^{10}$) this equation has the solution:
 \begin{equation} -t/b\tau_{ph}=x-x_0+\ln((x-1)/(x_0-1)),
\label{trelax}
\end{equation}
where $x_0=x(t=0)$ and $b\tau_{ph}$ is the spin-phonon relaxation
time ($T_{ph}=T_S\rightarrow T$). When the system is not far from
equilibrium ($x\sim1$), we get an exponential decay of the magnetization, 
with the same time constant $\tau_H=b\tau_{ph}$.  For a spin 1/2 system
\cite{A&B}:  \begin{equation} 
\tau_H=\alpha\frac{\tanh^2(\Delta_H/2k_BT)}{\Delta_H^2}, 
\label{tauform}
\end{equation}
with $\alpha=2\pi^2\hbar^2v^3N\tau_{ph}/3\Delta\omega$ ($N$ the molecule
density).

 The dynamical magnetization curves calculated in this model are given
Fig.~\ref{3Temps}b and Fig.~\ref{3rates}b. We started from equilibrium
($x_0=1$) in large negative fields. Then we let the system relax for a very
short time $\delta t$ and we calculated $x(\delta t)$ using Eq.~\ref{trelax}.
This value was taken as the initial value for the next field (the field step
is $r\delta t$). The parameters have been chosen to mimic the measured curves
of Fig.~\ref{3Temps}a and Fig.~\ref{3rates}a \cite{AD}. The obtained
similarity supports the possibility of the phonon bottleneck effect at the
timescale of a few 0.1 s. In the Fig.~\ref{3rates}a inset, we show the
variation of the calculated spin-phonon temperature $T_S$ for $T=0.1$~K and
$r=0.14$~T/s. After a cooling in negative fields, we can note a linear
variation in the plateau region (small positive fields), where $n_1/n_2\approx
cst.$. The slope of this quasi-adiabatic linear region gives the
plateau position and varies with the bath temperature and sweeping rate. 
In the Fig.~\ref{3rates}b inset we show the calculated field evolution of the
number of phonons at energy $\hbar\omega=\Delta_H$ at equilibrium
($T_{ph}=T_S=T$, dashed line) and out-of-equilibrium ($n_{T_{ph}}=n_{T=T_S}$,
$r=0.14$~T/s, black line). The difference between the two curves (thick segment
$\Delta\omega$) suggests the moving hole in the phonon distribution, while
their intersection gives the plateau intercept of the equilibrium
magnetization curve (above which the hole dissapears and $T_{ph}=T_S>T$). Let
note that in zero field the system is out-of-equilibrium even if magnetization
passes through the origin of coordinates (without a barrier, the switch
between $+1/2$ and $-1/2$ follows the level structure shown Fig.~\ref{3Temps}
inset ). At larger fields, in the plateau region, $n_1/n_2\approx cst.$ at
timescales shorter than $\tau_H=b\tau_{ph}$ (Eq.~\ref{tauform}), even after
the plateau crosses the equilibrium curve. Equilibrium is reached when
$\tau_H$ becomes small enough.

\begin{figure}
\begin{center}\includegraphics[width=8cm]{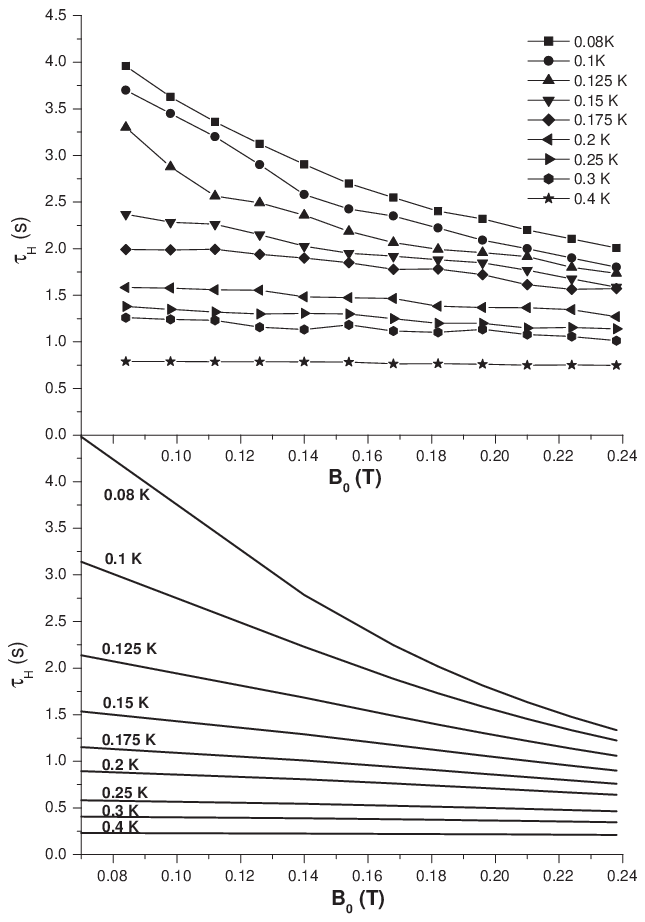}\end{center}

\caption{The relaxation times $\tau_H$, measured
(\textbf{a}$-$\emph{top}) and calculated (\textbf{b}$-$\emph{bottom}, same
parameters as in Fig.~\ref{3Temps}, \ref{3rates}b).}
 \label{taufit}
\end{figure}

Furthermore, we measured the relaxation of the magnetization of our crystal
at different fields and temperatures, along the plateau region. The
relaxation curves compared well to exponential decay and the obtained
relaxation times are presented in Fig.~\ref{taufit}a. The comparison with
those calculated (Fig.~\ref{taufit}b) is acceptable. But we noted that a
direct fit to Eq.~\ref{trelax} would necessitate larger values for $\alpha$ and
$\Delta_0$ ($\approx0.4-0.6$~sK$^2$ and $\approx0.2-0.3$~K).\\

\textbf{5. Conclusions}\\

In conclusion, this dynamical study of a single crystal of non-interacting
V$_{15}$ molecules with spin 1/2 is an example of a non-adiabatic
Landau-Zener model with dissipation. Due to relatively large splitting
$\Delta_0$ at the $S=1/2$ anticrossing in V$_{15}$ molecules (i) the
probability for ground-state adiabatic transitions is nearly equal to one and
(ii) the spin system absorbes phonons during the Landau-Zener transition
creating a hole in their distribution. The time and field evolution of this
hole generates a ``butterfly'' hysteresis loop quite different from the one of
high spin molecular magnets with large barrier and infinitesimal
$\Delta_0/k_BT$ ratio (no phonons are available at the anticrossing). The
effects presented in this paper seem to be a mark of molecules with low spin.

\acknowledgements
We are very pleased to thank P.C.E. Stamp, S. Myashita, I. Tupitsyn, A. K.
Zvezdin, H. De Raedt for useful discussions and E. Krickemeyer, P. K\"ogerler,
D. Mailly, C. Paulsen and J. Voiron for on-going collaborations.

\noindent $^{\sharp}$\emph{corresponding author: ichiores@polycnrs-gre.fr,
tel/fax: 00 33 4 76 88 11 94 / 11 91}

\begin{references}
\bibitem{hyp} For an isolated spin 1/2, $\Delta_0=0$, but different
couplings could generate a splitting, such as the hyperfine one
$A\vec{I}\cdot\vec{S}, I=7/2, S=1/2, A\approx10$~mK, $\Delta_0\approx10$~mK
\cite{Carter}. Also, the molecule symmetry allows the Dzyaloshinsky-Moriya
interactions that can generate $\Delta_0$ as large as 0.1~K \cite{Dobro}. A
value $\Delta_0\approx50$~mK is strongly supported by the experiment-to-model
comparison in Figs.~\ref{3Temps} and \ref{3rates}.
\bibitem{Stamp} The fast hyperfine fluctuations are characterized by the
transvers nuclear relaxation time $T_2$ associated with the dipolar
internuclear interactions and the total spread of the energy is related to the
hyperfine coupling A \cite{Stamp0}.
\bibitem{AD} $\alpha=0.15$~sK$^2$ ($2.9\cdot10^{-47}$~sJ$^2$), $A\approx10$~mK.
Taking $L\sim30-50~\mu m, N\sim 10^{27}$~m$^{-3},
\Delta\omega\sim5\cdot10^8$~s$^{-1}$, one gets a phonon velocity $v\approx
2800-3600$~m/s which is quite a reasonable value.
\end{references}
\end{document}